\documentclass{epl}

\usepackage[dvips]{epsfig,geometry}
\usepackage{amsmath}


\title{Rheology of distorted nematic liquid crystals}
\shorttitle{Rheology of distorted liquid crystals}
\author{D. Marenduzzo$^1$, E. Orlandini$^2$ and J.M. Yeomans$^1$}

\institute{$^1$Department of Physics, Theoretical Physics, 1 Keble Road,
  Oxford OX1 3NP, England\\
$^2$  INFM, Dipartimento di Fisica, Universita' di Padova, Via Marzolo 8, 
35131 Padova, Italy}

\begin{document}

\maketitle

\begin{abstract}
We use lattice Boltzmann simulations 
of the Beris--Edwards formulation of nematodynamics
to probe the response of a
nematic liquid crystal with conflicting anchoring at the boundaries under
shear and Poiseuille flow. The geometry we focus on is that of
the hybrid aligned nematic (HAN) cell, common in devices.
In the nematic phase, backflow effects resulting from
the elastic distortion in the director field render the velocity profile
strongly non-Newtonian and asymmetric. As
the transition to the isotropic phase is approached, 
these effects become progressively weaker. If the fluid is heated just 
above the transition point, however, another asymmetry appears,
in the dynamics of shear band formation. 
\end{abstract}


\section{Introduction}


Non-Newtonian behaviour in complex fluids such as liquid crystals 
arises because of the coupling between the molecular ordering,
described by the director field, and the 
flow field.  Liquid crystal molecules can be aligned by a shear or
rotational flow\cite{degennes,beris} and, conversely, movement of the molecules can induce
a velocity field, the 
so-called back-flow. This coupling results in flow-induced
instabilities, such as tumbling 
under shear, where the director field rotates in a
time-dependent steady state. 
An even more striking example is the
banded texture that is predicted upon shearing a nematic liquid
crystal in the vicinity of 
the nematic-isotropic transition\cite{olmsted,yuan}.
Moreover, in twisted nematic devices the phenomenon of optical 
bounce, a transient rotation of the
director field in the center of the cell in the wrong direction upon switching, 
has been well studied and is understood to be due solely to  
back-flow\cite{backflowtwist}. More recently,
there have been  
detailed measurements of back-flow effects on  
another commonly used cell, namely the hybrid
aligned nematic (HAN) cell \cite{hanbackflow}.

As a result of extensive theoretical, numerical and experimental
investigations the behaviour of nematic liquid crystals in 
shear and Poiseuille flow fields is well 
documented\cite{rey_review,shear_boundary1,shear_boundary2,
shear_boundary3,sluckin,nonalign1,nonalign2}. However as far as
we are aware results are limited to the cases where at {\em all}
the boundaries of the containing cell 
the alignment of the director field 
is either free, homogeneous (parallel
to the surface) or homeotropic (perpendicular to the surface).

This is perhaps surprising, given the 
widespread use in displays of nematics with
elastic distortions\cite{raynes,example1}. 
An elastic deformation appears in the unsheared state 
of a nematic liquid crystal when a rotation is
imposed on the director field by conflicting anchoring at 
opposite boundaries.
In trying to minimize its elastic free energy the system adopts a
uniform deformation across the sample. One might expect the 
interplay between such
a distorted director field and the flow field to lead to unexpected
non-Newtonian behaviour. Therefore the aim of this paper is
to present results for the behaviour of distorted
liquid crystalline systems under shear
and Poiseuille flow. 


We find that the imposed anchoring can strongly affect
both the transient and the
steady state properties of the flow. Focusing on a cell with 
mixed homeotropic and homogeneous anchoring, corresponding to the
geometry of the HAN cell,
we show that at low temperatures, deep in the nematic phase, the
elastic distortion  strongly affects the
velocity profile and introduces an asymmetry . 
As the temperature increases, or equivalently as the parameters change so as
to render director re-orientation due to the external flow
less costly, the asymmetry is progressively reduced.
As a result the liquid crystal rheology can, to a fair
approximation, be described by that of a Newtonian fluid.
Near the isotropic--nematic transition the
fluid again exhibits its non-Newtonian nature
and another asymmetry develops, this
time in the way banded textures develop in the system.

\section{Equations of motion}

We follow the Beris--Edwards formulation of nematodynamics\cite{beris} 
and write the hydrodynamic continuum equations of motion 
in terms of a tensor order
parameter ${\bf Q}$ whose largest eigenvalue, $\frac {2} {3} q$, $0<q<1$,
describes the magnitude of the order. 


The equilibrium properties of the liquid crystal are described 
by the following Landau-de Gennes free energy density 
\begin{equation}
f=\frac {A_0}{2} (1 - \frac {\gamma} {3}) Q_{\alpha \beta}^2 - 
          \frac {A_0 \gamma}{3} Q_{\alpha \beta}Q_{\beta
          \gamma}Q_{\gamma \alpha} 
+ \frac {A_0 \gamma}{4} (Q_{\alpha \beta}^2)^2 
+ \frac{K}{2} (\partial_\alpha Q_{\beta \gamma})^2,
\label{eqBulkFree}
\end{equation}
where $A_0$ is a constant and Greek indices label the Cartesian components 
of ${\bf Q}$. Eq. \ref{eqBulkFree} 
comprises a bulk contribution which describes a first order transition
from the isotropic to the uniaxial nematic 
phase at $\gamma=2.7$, together
with an elastic contribution -- penalizing distortions --
depending on the elastic constant $K$.

The anchoring of the director field on the boundary surfaces 
(Fig. 1) to a chosen director $\hat{n}^0$ is ensured by adding 
the surface term
\begin{equation}
f_{s}= \frac {1} {2} W_0 (Q_{\alpha \beta}-
                Q_{\alpha \beta}^0)^2, \qquad
Q_{\alpha \beta}^0=S_0(n^0_\alpha n^0_\beta-\delta_{\alpha \beta}/3).
\label{free_surface}
\end{equation}
The parameter $W_0$ controls the strength of the anchoring,
while $S_0$ determines the degree of surface order\cite{galatola,durand}.

The equation of motion for the nematic order parameter is \cite{beris}
\begin{equation}
(\partial_t+{\vec u}\cdot{\bf \nabla}){\bf Q}-{\bf S}({\bf W},{\bf
  Q})= \Gamma {\bf H}
\label{Qevolution}
\end{equation}
where $\Gamma$ is a collective rotational diffusion constant.
The term on the left-hand side is:
\begin{equation}\label{S_definition}
{\bf S}({\bf W},{\bf Q})
=(\xi{\bf D}+{\bf \Omega})({\bf Q}+{\bf I}/3)+({\bf Q}+
{\bf I}/3)(\xi{\bf D}-{\bf \Omega})
-2\xi({\bf Q}+{\bf I}/3){\mbox{Tr}}({\bf Q}{\bf W})
\end{equation}
where ${\bf D}=({\bf W}+{\bf W}^T)/2$ and
${\bf \Omega}=({\bf W}-{\bf W}^T)/2$
are the symmetric part and the anti-symmetric part respectively of the
velocity gradient tensor $W_{\alpha\beta}=\partial_\beta u_\alpha$.
The constant $\xi$ depends on the molecular
details of a given liquid crystal.
The molecular field ${\bf H}$
of Eq. (\ref{Qevolution})
describes the relaxation of the order parameter towards the minimum of
the free energy\cite{beris}. 

The fluid velocity, $\vec u$, obeys the continuity equation and
the Navier-Stokes equation, 
\begin{equation}
\rho(\partial_t+ u_\beta \partial_\beta)
u_\alpha = \partial_\beta \tau_{\alpha\beta}+\partial_\beta
\sigma_{\alpha\beta}+\eta \partial_\beta((1-3\partial_\rho
P_{0}) \partial_\gamma u_\gamma\delta_{\alpha\beta}+\partial_\alpha
u_\beta+\partial_\beta u_\alpha),
\label{navierstokes}
\end{equation}
where $\rho$ is the fluid density and $\eta$ is an isotropic
viscosity. 
The quantities $\sigma_{\alpha\beta}$ and $\tau_{\alpha\beta}$
are the symmetric and antisymmetric part of the stress tensor
respectively.

The equations for order parameter field Eq. (\ref{Qevolution}) 
and flow field Eq. (\ref{navierstokes}) are coupled. 
The velocity field and
its derivatives appear in the equation of motion for the order
parameter. 
Conversely, the order parameter field affects the dynamics of the flow field 
through the stress tensor 
which appears in the Navier-Stokes equation (\ref{navierstokes})
and depends on ${\bf Q}$ and ${\bf H}$. This back-action
of the order parameter field on the flow field is 
commonly termed backflow.  To solve these equations we use a lattice
Boltzmann algorithm, the details of which 
(for the two-dimensional case) have been given in
Ref.\cite{colin1} and further discussed in Ref. \cite{colin2}.

\section{Poiseuille flow in the nematic phase}

We consider a liquid crystal confined between two plates which lie 
parallel to the $x$--$y$ plane at $z=0$ and $z=L$. A rotation of 
$\pi/2$ about the $x$-axis, corresponding to a splay/bend deformation, 
is imposed on the director field by fixing the 
anchoring to be homogeneous along $\hat{{y}}$ at $z=L$ and
homeotropic (i.e. along $\hat{z}$) 
at $z=0$ as shown in Figure 1. The angle 
the director makes with the $z$ axis varies linearly 
between the plates. This geometry is the one defining
the HAN cell\cite{degennes,hanbackflow}.

\begin{figure}
\centerline{\psfig{figure=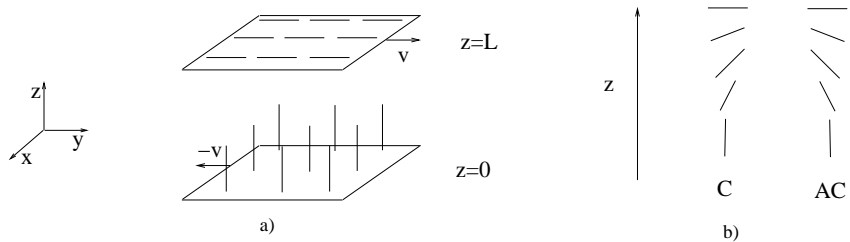,width=4.4in}}
\caption{a) Geometry used for the calculations described in the
text. The liquid crystal is sandwiched between two
infinite plates, parallel to the $xy$ plane, lying at
$z=0$ and $z=L$. The anchoring is homeotropic
on the bottom plate and homogeneous, along the flow direction,
on the top plate. The figure refers to the case of an imposed shear flow.
b) Sketch of the director field profile along the $\hat{z}$ direction
with no applied flow. The director can adopt the two 
degenerate configurations shown which we label 
as a clockwise (profile C) or anti-clockwise (AC) rotation.
Unless otherwise stated we focus here on the case in which
the director adopts the C configuration in the absence of flow.}
\end{figure}

We first report the results obtained when the
system is subject to a Poiseuille flow along $\hat{y}$. 
Figure 2a shows plots of the velocity field
across the system and the comparison to the parabolic profile,
expected for a Newtonian fluid, obtained if backflow is neglected.
(Backflow was 'turned off' in the numerical simulations
by imposing $\sigma_{\alpha\beta}=-P_0\delta_{\alpha\beta}$,
with $P_0$ a constant, and 
$\tau_{\alpha\beta}=0$ in Eq. (\ref{navierstokes}).) 
The difference between the two
cases is remarkable. Back-flow renders the rheology of the
liquid crystal markedly non-Newtonian.
First, the typical velocity 
in the system is more than a factor of two
smaller. 
Second, it is asymmetric.

It is possible to explain the asymmetry of the velocity 
in an intuitive way. 
The conflicting anchoring conditions of the HAN cell geometry 
yield a director profile in which a wide range of orientations 
coexist simultaneously in the sample in the absence of external perturbation 
(this remains true when the pressure difference 
applied to impose the flow 
is low). As these orientations couple differently to the imposed flow,
via the back-flow, the steady state displays an asymmetry.

\begin{figure}
\centerline{\psfig{figure=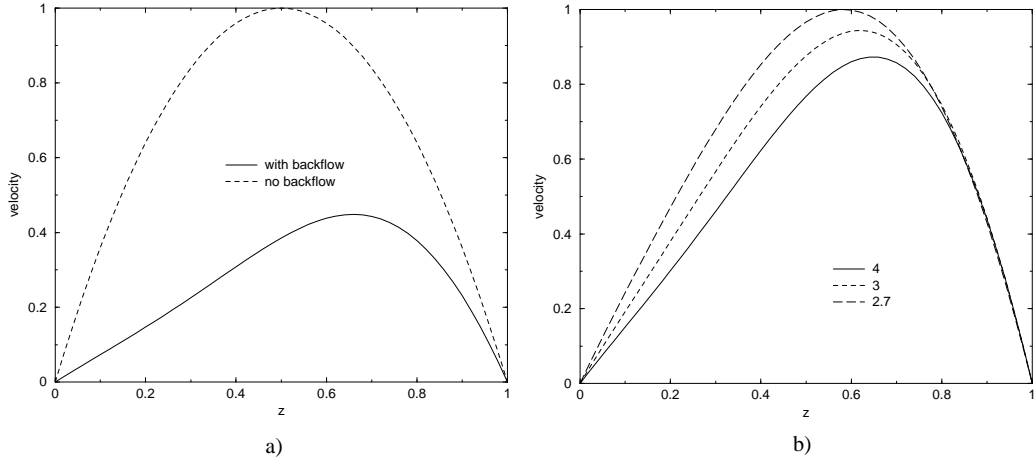,width=5.4in}}
\caption{Velocity profiles under Poiseuille flow.
(a) $\gamma=5$ for a channel of width 
$1$ $\mu$m. The velocity has been divided by 
$7$ $10^{-5}$ $\frac{A_0 L}{\Gamma}$
to make it dimensionless and with unit maximum.
(b) $\gamma=4,3,2.7$ for a channel of
width $1$ $\mu$m. The velocity has been divided by 
$3.8$ $10^{-5}$ $\frac{A_0 L}{\Gamma}$.}
\end{figure}

Figure 2b shows how the deviation from the 
parabolic velocity profile varies with 
temperature. (We consider $\gamma$ as an effective
temperature as it controls the isotropic-nematic transition.)
It can be seen that as the temperature
approaches that of the isotropic-nematic 
transition and thus the magnitude of the order decreases,
the velocity profile approaches that of a standard Newtonian
fluid.
Non-Newtonian effects also decrease when the system size is increased because
director re-orientation is easier and when the flow rate is increased because
the relative contribution of backflow is less.

\section{Shear flow in the nematic phase}

Let us now discuss the behaviour of the same HAN cell under a
constant applied shear.
Consider first an imposed shear along $\hat{{y}}$ (Fig. 1). 
In a Newtonian fluid this would give a linear velocity profile. 
However now, in
steady state flow conditions, the strain rate is not constant,
and the velocity field deviates from linear because of the 
different coupling of the flow to the director field as ${{z}}$ varies.

Fig. 3 shows how the scaled velocity profile in the channel 
depends on the system parameters.
When director field reorientation is more difficult
due to a narrow width (Fig. 3a), to a low temperature 
(data not shown) or to low shear rates (Fig. 3b)
the asymmetry is more pronounced. This effect is qualitatively
similar to the one observed with Poiseuille flow.

Consider the case of increasing shear rate (Fig. 3b)
when an interesting structure appears in the director field
as the fluid velocity increases. The cell tends to spontaneously
separate into two distinct regions characterised by different director 
configurations with a rather sharp crossover between them.
In the top portion of the cell, the strain rate is higher and the fluid 
is well aligned with the flow and resembles a standard nematic.
All the elastic distortion is confined to the bottom
portion of the cell where the strain rate gradually decreases.
The boundary between the two regions moves nearer to the 
homeotropic plate as the shear rate increases.
This is an example of shear-induced patterning of the director configuration
across the cell, a phenomenon which we will encounter again below, 
when we discuss banding dynamics near the isotropic-nematic transition.

\begin{figure}
\centerline{\psfig{figure=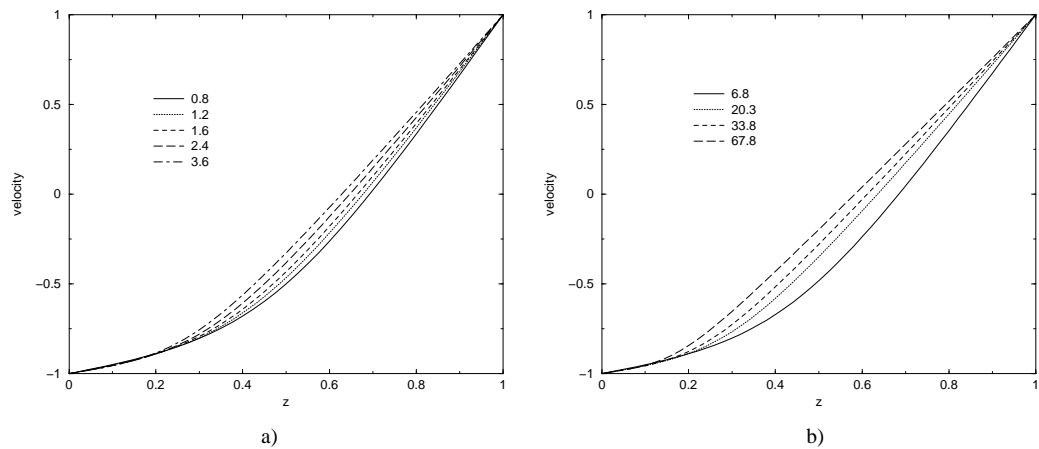,width=5.4in}}
\caption{Velocity profiles across the cell under shear flow
for different values of the model parameters. In the plots $z$ is scaled
by the system dimension and the velocity is scaled by the
maximum velocity in the system, attained at $z=L$.
The elastic constant $K=11.7$ pN for all plots.
(a) Velocity profiles for different cell widths $L$ 
(in micrometers). The shearing velocity
is constant ($0.7$ mm s$^{-1}$),
and $\gamma=5$ for these plots.
(b) Velocity profiles for a cell of width
$1$ $\mu$m, with $\gamma=5$ and variable shearing velocities 
(in dimensionless units, scaled by $10^5$ $\frac{A_0 L}{\Gamma}$).}
\end{figure}
 
A few remarks are now in order: 
first, it is to be noted that the two configurations in
which the director performs a clockwise or anti-clockwise rotation 
around the $\hat{x}$ direction (configurations C and AC in Fig.1), 
which are equally probable in the absence of flow,
couple differently to the imposed flow. The results we have just shown  
refer to the case of a clockwise rotation. 
If the starting configuration is that labelled AC in Fig. 1, 
the details of the flow change.
In a region which lies at the bottom of the cell and which
increases in  size with increasing shear, the director field closely resembles
that of a smaller cell with homeotropic anchoring 
on both plates which is subjected to a shear flow.

Second, the behaviour of a nematic liquid crystal in an imposed flow
can be either flow aligning or flow tumbling\cite{rey_review}.
In the Beris--Edwards 
model, the behaviour depends on the magnitude of the order and on 
the parameter $\xi$\cite{beris,colin1}. We chose $\xi=0.85$ 
so that the liquid crystal is flow aligning
in the whole temperature ($\gamma$) regime considered.
Flow tumbling materials will be considered elsewhere.

Third, we have so far considered the case in which the director
field is initially in the flow plane.
If the system is sheared along the 
$\hat{{x}}$ direction instead, a small secondary flow 
develops along $\hat{{y}}$. 

\section{Shear banding near the nematic--isotropic transition}

When a nematic is sheared in the vicinity of the isotropic--nematic
boundary (just above the transition temperature) 
shear bands can form. These are coexisting states, one
nematic, one paranematic, with an interface between them which
normally lines up along the flow direction. The amount of each
coexisting phase depends on the shear rate and the free energies of
the competing phases\cite{olmsted} in a way that is not yet fully understood.

When free (i.e. unconstrained) 
boundary conditions are considered\cite{olmsted}, 
two ordered bands,
in which the fluid is locally in a nematic state, 
can form at the surface of the cell,
because of the favourable coupling of an ordered phase to 
the higher strain rates present near the moving surfaces.
These grow symmetrically until they reach a steady state width.

Here we consider the formation of the shear bands in a
HAN cell geometry (Figure 1) 
and show that  the conflicting boundary conditions
introduce an asymmetry into the kinetic pathway by which the
bands form. We work at
$\gamma=2.65$ and surface order
$S_0$ between $0.05$ and $0.15$ 
so that the fluid is in the isotropic
phase, and apply a shear of a few mm s$^{-1}$ (Fig. 4). 
Because $S_0$ and $W_0$ are both non-zero, 
two thin layers of paranematic order,
of thickness a few tens of nm,
appear near the boundaries even in the
unperturbed state \cite{galatola,durand}.

The dynamical evolution of the liquid crystal 
once shear is applied, in the case of conflicting
anchoring at the boundaries, is shown in Fig. 4.  
First, two bands form at the edges of the cell
(as is the case for free boundaries).
This is because, as in realistic situations, there is initially
a transient large strain rate at the boundaries as it takes time
for the shear perturbation to reach the bulk.
Quite rapidly, however (after a few ms according to our simulations), 
the  band near the homeotropic surface becomes unstable because 
the surface 
anchoring conflicts with the flow alignment angle favoured by the
presence of the shear. 
Thus, the band quickly unbinds 
from the surface and becomes thinner. 
It moves across the system to join the band forming at the other surface. 
This is advantageous because it minimizes
the number of interfaces. 

At later times, the final steady state reached in the
simulations depends on the degree and direction of surface ordering
in the top plate. For the case displayed in Fig. 4 
($S_0=0.08$ and homogeneous anchoring along the flow direction),
the wider band finally unbinds from the top surface
and eventually comes to rest in the centre of the system.
When the degree of surface ordering is weaker, 
the stable steady state corresponds to the
band remaining at the top of the sample.
Mapping the simulation parameters onto physical parameters we
estimate that the evolution of the bands takes a few seconds, 
in a cell of width $\sim$$\mu$m, a time scale 
easily amenable to experiments.

\begin{figure}
\centerline{\psfig{figure=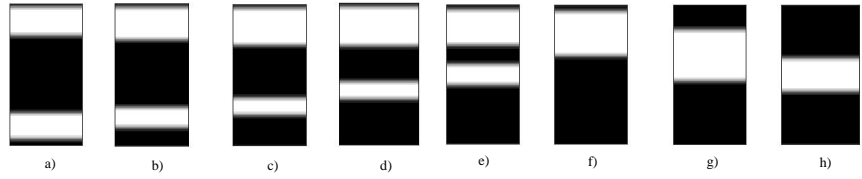,width=4.5in}}
\caption{Pictures of the formation of 
shear bands when the
boundaries are anchored as described in the text. In this calculation
we put the anchoring strength $S_0=0.08$ and $\gamma=2.65$,
while the dimensionless ratio $\frac{\Gamma v}{A_0L}=7.35$ $10^{-4}$.
The plots show the magnitude of the order parameter.
Black regions correspond to low order (the system is locally isotropic
with $q$ very small)
and white regions to high order (the system is locally nematic
with $q\sim 0.3$). 
The graphs correspond to times 
(a) 0.9 ms, (b) 49.5 ms, (c) 119.7 ms,
(d) 270 ms, (e) 711 ms, (f) 855 ms, (g) 911.7 ms, (h) 5.4 s.}
\end{figure}

\section{Discussion}

In conclusion, we have studied the response of a nematic
liquid crystal with conflicting anchoring at the boundaries,
and hence an imposed director distortion, 
under shear and Poiseuille flow. 
Specifically, we focussed on the geometry
of the hybrid aligned nematic cell and 
found that when the system is nematic and the strain rate is low, 
the imposed distortion has a strong effect 
on the velocity profile via backflow.
As a result the fluid behaves in a markedly non-Newtonian
manner. 

For Poiseuille flow, the effect is particularly visible. The velocity profile 
becomes strongly asymmetric and the flow rate is
strongly suppressed. 
Under shear, as the shear rate increases, the
velocity profile forms
two distinct regions with different typical
strain rates.
One of these regions, which contains virtually all the director deformation, 
is increasingly confined near one of the boundaries of the cell.

As the degree of order in the sample decreases, due for example to
an increase of temperature towards the transition temperature to the
isotropic phase, the strain rates across the system 
become more symmetric and
the fluid rheology resembles that of a Newtonian fluid.
However, just above the transition temperature,
we again observe non-Newtonian
behaviour, namely shear banding.
The conflicting
surface-induced ordering affects the way in which
the bands form, introducing a dynamical asymmetry.

A lattice Boltzmann solution of the
Beris--Edwards equations proved very powerful for this study.
It allowed us to consider fully the effects of temperature, 
backflow, conflicting anchoring and different
degrees of order within the cell on the liquid crystal rheology.

There are several directions in which this work could be pursued.
First, one can extend the present treatment to 
flow tumbling materials and to other cells with elastic distortions,
for example the twisted nematic cell, a basic display device.
Second, it would be very interesting
to study flow in the cells with more complicated boundary conditions which 
are increasingly being fabricated by modern lithographic techniques.
Third, we should like to compare the present picture to the one 
arising from rheological studies in which distortions are not
imposed but are natural, as is the case for example in cholesteric 
liquid crystals.

Finally, we note that the phenomenology described here 
is robust over a wide range of parameters in our calculations,
and occurs on physically accessible
length and time scales.
The results can be tested experimentally with technology
currently available.

{\em Acknowledgements:} We thank C. Denniston for 
useful and important discussions. 


\end{document}